\documentclass[%
reprint,
superscriptaddress,
amsmath,amssymb,
longbibliography,
aps,
]{revtex4-1}

\usepackage{hyperref}
\usepackage{graphicx}%
\usepackage{verbatim}
\usepackage[version=4]{mhchem}
\usepackage[dvipsnames]{xcolor}
\usepackage{physics}
\usepackage[normalem]{ulem}
\usepackage{pdfpages}
\usepackage{pgffor}
\usepackage{multirow}
\usepackage{bm}
\usepackage{booktabs}
\usepackage{xr}

\newif\ifhighlight

\newcommand{\highlight}{\highlighttrue}
\highlight
\newcommand{\editor}[2]{%
  \expandafter\newcommand\csname #1note\endcsname[1]{%
    \textcolor{#2}{(\textbf{#1:} ##1)}}%
  \expandafter\newcommand\csname #1\endcsname[1]{%
    \ifhighlight\textcolor{#2}{##1} \else ##1\fi}%
  \expandafter\newcommand\csname #1cancel\endcsname[1]{%
    \ifhighlight\textcolor{#2}{\sout{##1}}\fi}%
  \expandafter\newcommand\csname #1change\endcsname[2]{%
    \ifhighlight\textcolor{#2}{\sout{##1} ##2}\else ##2\fi}%
  \newenvironment{#1text}{\ifhighlight\color{#2}\fi}{\color{black}}
}

\editor{GKC}{brown}
\editor{PP}{Green}
\editor{DS}{Red}
\editor{JN}{blue}
\editor{HT}{orange}
\editor{MC}{teal}
\editor{SSa}{purple} 

\makeatletter
\AtBeginDocument{\let\LS@rot\@undefined}
\makeatother

\ifdefined\DIRACREP
\else
\newcommand{\DIRACREP}{}

\usepackage{physics}
\usepackage{xparse}
\ifdefined\COSMOMATHS
\else
\newcommand{\COSMOMATHS}{}

\newcommand{\mbf}[1]{\ensuremath{\mathbf{#1}}}

\fi %

\NewDocumentCommand{\rep}{s d<| d|>}{%
\IfBooleanTF{#1}{
   \IfValueTF{#2}{
       \IfValueTF{#3}{\braket{#2}{#3}}{\bra{#2}}
       }{
       \IfValueTF{#3}{\ket{#3}}{}
       }
   }{
   \IfValueTF{#2}{
       \IfValueTF{#3}{\braket*{#2}{#3}}{\bra*{#2}}
       }{
       \IfValueTF{#3}{\ket*{#3}}{}
       }
   }
}

\NewDocumentCommand{\rbra}{sm}{\IfBooleanTF{#1}{\rep*<#2|}{\rep<#2|}}
\NewDocumentCommand{\rket}{sm}{\IfBooleanTF{#1}{\rep*|#2>}{\rep|#2>}}
\NewDocumentCommand{\rbraket}{smom}{
    \IfBooleanTF{#1}{
        \IfNoValueTF{#3}{\rep*<#2||#4>}{\rep*<#2|#3\rep*|#4>}
    }{
        \IfNoValueTF{#3}{\rep<#2||#4>}{\rep<#2|#3\rep|#4>}
    }
}

\NewDocumentCommand{\cg}{m m m}{\rep<#1; #2||#3>}

\NewDocumentCommand{\field}{o m e{_} e{^} o e{_} e{^}}{
\IfValueTF{#5}{\overline{
  #2\IfValueT{#3}{_#3}\IfValueT{#4}{^{\otimes #4}} %
  \otimes
  #5\IfValueT{#6}{_#6}\IfValueT{#7}{^{\otimes #7}} %
  \IfValueT{#1}{;#1}
}}{
  \IfValueTF{#4}{\overline{
     #2\IfValueT{#3}{_#3}\IfValueT{#4}{^{\otimes #4}}
     \IfValueT{#1}{;#1}
  }}
  {#2\IfValueT{#3}{_#3}}
}
}

\NewDocumentCommand{\frho}{o e{_} e{^}}{
\field[#1]{\rho}_{#2}^{#3}
}

\newcommand{\bx}{\mbf{x}}
\newcommand{\bxhat}{\hat{\mbf{x}}}

\newcommand{\e}{a}  %

\NewDocumentCommand{\ex}{e_}{
\IfValueTF{#1}{\e_{#1}\bx_{#1}}{\e\bx}
}  %

\NewDocumentCommand{\lm}{e_}{
\IfValueTF{#1}{l_{#1}m_{#1}}{lm}
}
\NewDocumentCommand{\nlm}{e_}{
\IfValueTF{#1}{n_{#1}\lm_{#1}}{n\lm}
}
\NewDocumentCommand{\enlm}{e_}{
\IfValueTF{#1}{\e_{#1}\nlm_{#1}}{\e\nlm}
}
\NewDocumentCommand{\en}{e_}{
\IfValueTF{#1}{\e_{#1}n_{#1}}{\e n}
}
\NewDocumentCommand{\nlk}{e_}{
\IfValueTF{#1}{n_{#1}l_{#1}k_{#1}}{nlk}
}
\NewDocumentCommand{\enlk}{e_}{
\IfValueTF{#1}{\e_{#1}\nlk_{#1}}{\e\nlk}
}
\NewDocumentCommand{\enl}{e_}{
\IfValueTF{#1}{\en_{#1}l_#1}{\en l}
}

\NewDocumentCommand{\nnl}{s}{
\IfBooleanTF{#1}{n_1 n_2 l}{n_1; n_2; l}
}
\NewDocumentCommand{\ennl}{s}{
\IfBooleanTF{#1}{\en_1 \en_2 l}{\en_1; \en_2; l}
}

\NewDocumentCommand{\gslm}{s}{
\IfBooleanTF{#1}{\sigma\lambda\mu}{\sigma;\lambda\mu}
}
\NewDocumentCommand{\glm}{}{\lambda\mu}

\fi %

\ifdefined\COSMOMATHS
\else
\newcommand{\COSMOMATHS}{}

\newcommand{\mbf}[1]{\ensuremath{\mathbf{#1}}}

\fi %

\definecolor{ForestGreen}{RGB}{34,139,34}

\makeatletter
\let\oldket\ket 
\def\ket{\@ifstar{\oldket}{\oldket*}}
\let\oldbra\bra
\def\bra{\@ifstar{\oldbra}{\oldbra*}}
\let\oldev\ev
\def\ev{\@ifstar{\oldev}{\oldev*}}
\makeatother

\usepackage{xparse}
\NewDocumentCommand{\fket}{s m}{ \IfBooleanTF{#1}{\oldket{#2}}{\oldket*{#2}}}
\usepackage{xparse}
\NewDocumentCommand{\fbra}{s m}{ \IfBooleanTF{#1}{\oldbra{#2}}{\oldbra*{#2}}}

\usepackage{physics}
\usepackage{xparse}

\NewDocumentCommand\te{s}{\tilde{\e}\IfBooleanTF{#1}{'}{}}
\NewDocumentCommand\tn{s}{\tilde{n}\IfBooleanTF{#1}{'}{}}
\NewDocumentCommand\tl{s}{\tilde{l}\IfBooleanTF{#1}{'}{}}
\NewDocumentCommand\tm{s}{\tilde{m}\IfBooleanTF{#1}{'}{}}
\NewDocumentCommand\tlm{s}{\IfBooleanTF{#1}{\tl*\tm*}{\tl\tm}}
\NewDocumentCommand\tnlm{s}{\IfBooleanTF{#1}{\tnl*\tm*}{\tnl\tm}}
\NewDocumentCommand\tnl{s}{\IfBooleanTF{#1}{\tn*\tl*}{\tn\tl}}

\NewDocumentCommand\ttn{s}{\tilde{N}\IfBooleanTF{#1}{'}{}}
\NewDocumentCommand\tttl{s}{\tilde{L}\IfBooleanTF{#1}{'}{}}
\NewDocumentCommand\ttm{s}{\tilde{M}\IfBooleanTF{#1}{'}{}}
\NewDocumentCommand\ttlm{s}{\IfBooleanTF{#1}{\tttl*\tm*}{\tttl\tm}}
\NewDocumentCommand\ttnlm{s}{\IfBooleanTF{#1}{\ttnl*\ttm*}{\ttnl\ttm}}
\NewDocumentCommand\ttnl{s}{\IfBooleanTF{#1}{\ttn*\tttl*}{\ttn\tttl}}

\usepackage{bm,upgreek}

\usepackage{color}

\newcommand{\ds}[1]{\color{red}}
\begin{document}

\title{Exploring the design space of machine-learning models \\for quantum chemistry with a fully differentiable framework}
\author{Divya Suman}
\altaffiliation{These authors contributed equally to this work}
\affiliation{Laboratory of Computational Science and Modeling, Institut des Mat\'eriaux, \'Ecole Polytechnique F\'ed\'erale de Lausanne, 1015 Lausanne, Switzerland}

\author{Jigyasa Nigam}
\altaffiliation{These authors contributed equally to this work}
\email{jnigam@mit.edu}
\affiliation{Laboratory of Computational Science and Modeling, Institut des Mat\'eriaux, \'Ecole Polytechnique F\'ed\'erale de Lausanne, 1015 Lausanne, Switzerland}

\author{Sandra Saade}
\affiliation{Laboratory of Computational Science and Modeling, Institut des Mat\'eriaux, \'Ecole Polytechnique F\'ed\'erale de Lausanne, 1015 Lausanne, Switzerland}

\author{Paolo Pegolo}
\affiliation{Laboratory of Computational Science and Modeling, Institut des Mat\'eriaux, \'Ecole Polytechnique F\'ed\'erale de Lausanne, 1015 Lausanne, Switzerland}

\author{Hanna T\"urk}
\affiliation{Laboratory of Computational Science and Modeling, Institut des Mat\'eriaux, \'Ecole Polytechnique F\'ed\'erale de Lausanne, 1015 Lausanne, Switzerland}

\author{Xing Zhang}
\affiliation{Division of Chemistry and Chemical Engineering, California Institute of Technology, Pasadena, CA 91125, USA}
\author{Garnet Kin-Lic Chan}
\affiliation{Division of Chemistry and Chemical Engineering, California Institute of Technology, Pasadena, CA 91125, USA}

\author{Michele Ceriotti}
\email{michele.ceriotti@epfl.ch}
\affiliation{Laboratory of Computational Science and Modeling, Institut des Mat\'eriaux, \'Ecole Polytechnique F\'ed\'erale de Lausanne, 1015 Lausanne, Switzerland}
\affiliation{Division of Chemistry and Chemical Engineering, California Institute of Technology, Pasadena, CA 91125, USA}

\date{\today}%

\begin{abstract}
Traditional atomistic machine learning (ML) models serve as surrogates for quantum mechanical (QM) properties, predicting quantities such as dipole moments and polarizabilities, directly from compositions and geometries of atomic configurations. 
With the emergence of ML approaches to predict the ``ingredients'' of a QM calculation, such as the ground state charge density or the effective single-particle Hamiltonian, it has become possible to obtain multiple properties through analytical physics-based operations on these intermediate ML predictions. 
We present a framework to seamlessly integrate the prediction of an effective electronic Hamiltonian, for both molecular and condensed-phase systems, with \textsc{PySCFAD}, a differentiable QM workflow that facilitates its 
indirect training against functions of the Hamiltonian, such as electronic energy levels, dipole moments, polarizability, etc. %
We then use this framework to explore various possible choices within the design space of hybrid ML/QM models, examining the influence of incorporating multiple targets on model performance and learning a reduced-basis ML Hamiltonian that can reproduce targets computed from a much larger basis. Our benchmarks evaluate the accuracy and transferability of these hybrid models, compare them against predictions of atomic properties from their surrogate models, and provide indications to guide the design of the interface between the ML and QM components of the model.

\end{abstract}

\maketitle

\section{Introduction}

Machine learning (ML) has become indispensable for atomistic modeling of molecules and materials, driving scientific discovery and accelerating the search for compounds with distinct properties. ML approaches have not only enabled large-scale molecular dynamics through accurate predictions of potential energy surfaces~\cite{behl-parr07prl,gap2013, SNAP, nequip, mace,bowm+09jcp,rupp+12prl, drau19prb, behler2021four}, but also refined our understanding of the intricate relationships between atomic geometries and physical observables, including electronic properties such as dipole moments~\cite{veit+20jcp, sun2022molecular} and polarizabilities~\cite{wilk+19pnas,zhan+20prb,zhang2023universal, feng2023accurate}.

Although these surrogate models can describe complex molecular behaviors, they often cannot infer properties beyond those they were trained to reproduce, or struggle to extrapolate to structures that deviate considerably from those included in the training set~\cite{wilk+19pnas, veit+20jcp}. 
One of the possible approaches to address these limitations is to develop models that aim for broader transferability by learning fundamental quantities central to the electronic structure problem, such as the electron density~\cite{grisafi2018transferable, lewis2021learning,fabr+20jcp, rackers2023recipe,koker2024higher,li2024image} or $N$-electron density matrices~\cite{shao2023machine, febrer2024graph2mat, zhang2025symmetrypreservingtransferablerepresentationlearning}, the electronic wavefunction~\cite{pfau2020ab, schatzle2023deepqmc, hermann2023ab, Pescia2024, scherbela2024towards,gerard2024transferable} or representations of an effective single-particle Hamiltonian operator in a specified atomic orbital (AO) basis~\cite{schu+19nc, li2022deeph, gong2023natcom, zhang2022equivariant}.
Learning any of these underlying electronic quantities, at a given level of theory, grants access, in principle, to all observables that can be obtained from relatively inexpensive postprocessing operations. %
In this work, we focus on the problem of learning an effective single-particle Hamiltonian.

The matrix elements of the Hamiltonian $\mathbf{H}$ encode pairwise interactions between the AO basis functions, which can be centered on either the same atom (constituting on-site interactions) or two different atoms (off-site interactions) in a given molecular configuration. 
A key property of $\mathbf{H}$ is its equivariance under operations of the O(3) group of rotations and inversions,  and permutations of identical atoms~\cite{niga+22jcp}. 
Early ML approaches for modeling $\mathbf{H}$ often circumvented explicitly describing these symmetries by using data augmentation~\cite{schu+19nc}, relying on ad-hoc modifications of atom-centered descriptors~\cite{hegde2017machine}, or by targeting observables such as optical excitations~\cite{west-maur21cs} through an intermediate \emph{invariant} representation of the Hamiltonian with respect to geometric transformations.
More recent approaches handle symmetries directly and account for pair dependence of matrix elements using bond-centered atomic cluster expansion descriptors~\cite{zhang2022equivariant} or through equivariant features for the corresponding atom pairs~\cite{niga+22jcp}. Equivariant approaches based on message-passing neural networks~\cite{unke2021se3equivariant, gong2023natcom,yin2024towards, wang2024deeph2,zhong2023transferable,zhong2024universal} have similarly risen to the task, as their underlying equivariant node and edge features can be naturally adapted to learn pairwise quantities.

Within a framework for Hamiltonian machine learning, there are several possibilities to define the target. For instance, one may learn the exact matrix representation corresponding to a mean-field calculation performed using a specified electronic structure method and basis set, accepting the fact that any truncated basis will introduce a finite-basis set error. An alternative would be to learn a reduced effective $\mathbf{H}$ that reproduces observables from a higher-level of theory or a larger-basis calculation.
Using this strategy of integrating ML and quantum chemistry, Ref.~\citenum{cignoni2024electronic} optimizes an effective minimal-basis $\mathbf{H}$ to match properties (eigenspectrum, L\"owdin charges) derived from quantum mechanical (QM) calculations performed on a much larger basis set. A similar framework~\cite{tang2024approaching} refines an effective one-electron Hamiltonian by machine learning a correction, so that multiple properties derived from it align with reference data computed using higher-accuracy many-body perturbative methods such as CCSD. Such hybrid approaches have demonstrated improved accuracy and transferability of ML at a lower computational cost. 
Interfacing ML approaches with quantum chemistry is made simpler by the incorporation of automatic differentiation (AD) capabilities within electronic structure codes. 
While this effort is useful in its own right, as it allows one to compute properties that were previously inaccessible due to the absence or impracticality of analytical derivatives, it is especially useful in combination with ML frameworks designed around evaluating gradients of a loss function through backpropagation, and has been explored in several recent works~\cite{li-yaron2018, tan2023automatic,stishenko2024integrated}. 
By training ML models on physical quantities derived from electronic structure calculations in a fully differentiable framework, one can attempt to improve predictive accuracy and enhance transferability across chemical systems. 

Developing effective ML models for intermediate descriptions of the electronic structure requires accounting for the subtleties of quantum chemistry (including the choice of the level of theory and basis set) in addition to conventional ML tasks (such as selecting appropriate architectures).
In this work, we combine our previously-proposed hybrid (or indirect) ML architecture~\cite{cignoni2024electronic} with the auto-differentiable electronic structure code \textsc{PySCFAD}~\cite{zhang2022differentiable} to optimize an intermediate representation of $\mathbf{H}$ that reproduces target electronic properties computed by explicit QM manipulation of a machine-learned effective Hamiltonian. 
In particular, we target molecular dipole moments ($\bm{\mu}$) and polarizabilities ($\bm{\alpha}$), as well as band energies in a condensed phase system. 
We keep a minimalistic symmetry-adapted parameterization of $\mathbf{H}$, and systematically explore how variations in quantum chemical design choices affect the performance of the model by progressively increasing the physical constraints in our model trained on a diverse subset of the QM7 dataset~\cite{rupp2012fast,blum,montavon2013machine} and evaluating its ability to generalize to larger systems, including QM9~\cite{ramakrishnan2014quantum, ruddig}.
Our results demonstrate that a well-constrained indirect model often improves predictive accuracy for both QM7 and QM9 datasets, especially for response properties such as polarizability. When trained to reproduce properties from larger-basis set calculations, the effective indirect minimal-basis models achieve an accuracy comparable to that of their large-basis counterparts. while remaining more computationally efficient. 
Even if the effective indirect minimal-basis models are less accurate when trained against properties from large-basis set calculations than when trained against a consistent level of theory, their accuracy is much better than the basis set error. We show that it is, therefore, preferable to train against large-basis outputs, as the predictions are more accurate in an absolute sense, even though the model has the size and computational cost of a minimal basis model.

We also evaluate their transferability on larger, more complex molecules than the training set, where the property-specific surrogate models have previously been limited due to their inability to capture non-local effects~\cite{veit+20jcp}. 
Our approach contributes to the broader effort of integrating ML with QM, with models that combine the interpretability and transferability of physics-based approaches with the efficiency of data-driven methods.

\begin{figure*}[tb]
\centering
\includegraphics[width=0.8\linewidth]{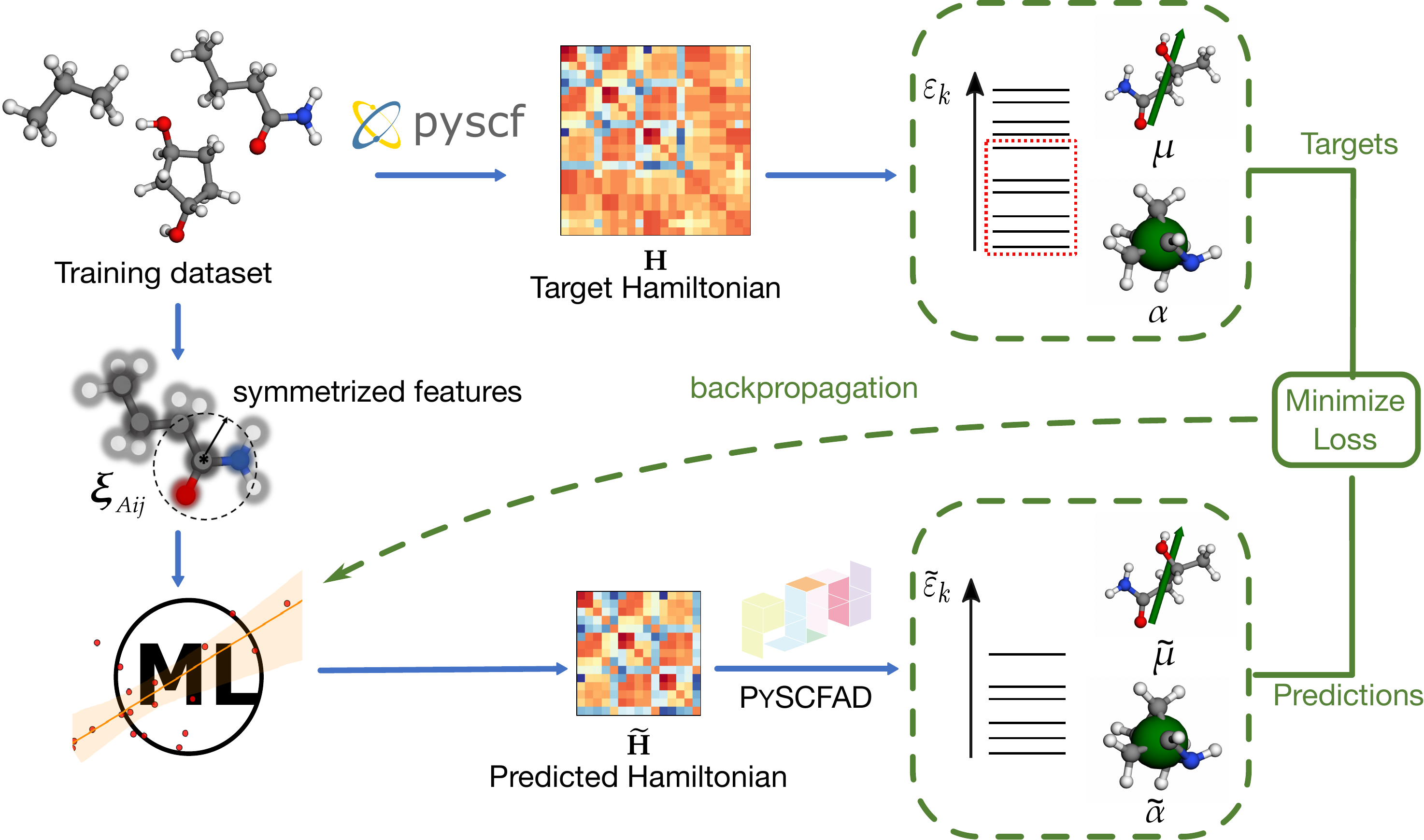}
\caption{Schematic workflow depicting the modular integration of the ML prediction of electronic Hamiltonians (for a selected basis set) with \textsc{PySCFAD}. The upper branch of the workflow demonstrates the generation of the reference Hamiltonian corresponding to an atomic configuration through PySCF, from which observables including the dipole moment ($\bm{\mu}$), polarizability ($\bm{\alpha}$), and MO energies ($\varepsilon_k$) are computed. 
These quantities serve as the targets for various models described in the text. Symmetry-adapted pair features $\xi_{A_{ij}}$ computed for each pair of atoms ($i,j$) in each molecular configuration ($A$) are used as inputs to an ML model that yields a prediction of the Hamiltonian. When interfaced with \textsc{PySCFAD}, these predictions can be used to compute the corresponding predictions of molecular properties ($\Tilde{\bm{\mu}}, \Tilde{\bm{\alpha}}, \Tilde{\varepsilon_k}$). The model can be optimized to minimize the loss directly on the Hamiltonian prediction or, alternatively, minimize the loss on the secondary properties derived from it.}
    \label{fig:model-schematic}
\end{figure*}

\section{Theory and methods}

In many quantum chemistry frameworks, the effective single-electron wavefunction is expanded in terms of localized orbital basis functions centered on atomic positions (atomic orbitals, AO). 
These basis functions are typically non-orthogonal due to their localized nature. The single-electron eigenstates are obtained by solving a generalized eigenvalue equation including the overlap matrix between the basis functions, $\mathbf{S}$, 
\begin{equation}\label{eq:eigv-gen}
    \mathbf{H} \mathbf{C} = \mathbf{S} \mathbf{C}\operatorname{diag}\boldsymbol{\varepsilon}
\end{equation}
where $\bm{\varepsilon}$ is the vector of one-electron eigenvalues, or molecular orbital (MO) energies, and $\mathbf{C}$ is the (unitary) matrix of MO coefficients. 
In mean-field theories, $\mathbf{H}$ depends self-consistently on the MOs, but in the following, we will assume that $\mathbf{C}$ can be obtained from a single diagonalization.
The one-electron density matrix $\bm{\rho}$ is then computed from the MO coefficients and the occupation numbers $\mathbf{f}$ of the molecular orbitals as,
\begin{align}\label{eq: rho}
    \bm{\rho} = \mathbf{C} \operatorname{diag}(\mathbf{f}) \mathbf{C}^\dagger.
\end{align} 
Any ground state property can then be computed from the MO energies and the density matrix. For instance, the total dipole moment $\bm{\mu}$ of an atomic structure $A$ can be computed as
\begin{equation}\label{eq:dipole-moment}
    \bm{\mu}(A) = - e \Tr(\bm{\rho}_A \mathbf{x}_A) + \sum_{i \in A} Q_i \mathbf{r}_i.
\end{equation}
The first term is the electronic contribution to the dipole moment, where $e$ is the electronic charge and $\mathbf{x}_A$ is the representation of the electronic position operator $\hat{\mathbf{x}}$ in the AO basis of structure $A$, while the second term is the nuclear contribution, computed as the product of the effective nuclear charges, $Q_i$, and positions, $\mathbf{r}_i$, of each atom $i$.
Response properties, in turn, are obtained in the linear approximation as the derivatives of ground state observables.
The polarizability tensor $\bm{\alpha}$, for instance, can be computed as the (zero-field) derivative of the dipole with respect to an applied electric field.
Thus, modeling the Hamiltonian matrix $\mathbf{H}$ enables the prediction of a wide range of electronic properties. 

It is possible to train ML models to directly predict each property starting from atomic species and coordinates, without explicitly modeling the Hamiltonian. This scheme, however, requires a separate ML model to be trained for each property of interest. 
We use these direct property models as a baseline for comparison with effective Hamiltonian models that provide a physically interpretable framework from which various properties can be derived simultaneously. As we shall see, this second approach also offers better transferability to out-of-sample molecules. 
As explained below, we use simple linear models, even though they are not especially accurate, in order to focus our attention on the role played by the many choices one can make in the design of the part of the model that more closely follows the structure of explicit QM calculations.

\subsection{ML models of atomistic properties}\label{sec:property_models}

The dipole moment is usually represented as a Cartesian tensor of rank one, whereas polarizability is a Cartesian tensor of rank two. Although it is possible to model these properties in this form, it is usually more convenient to decompose them into irreducible representations (irreps) of O(3), each of which can be individually modeled, as in Refs.~\onlinecite{wilk+19pnas, veit+20jcp},
\begin{equation}
    y_{A} \equiv \bigoplus_{\sigma \lambda} y_{A}^{\sigma \lambda}.
\end{equation}
In other words, the Cartesian representation of the property $y$ for structure $A$ ($y_{A}$) is expressed as a direct sum of O(3) irreps indexed by $\sigma \lambda$, each with a behavior under spatial inversion denoted by $\sigma$, and rotational behavior as a spherical harmonic $Y_{\lambda}^\mu(\bxhat)$ which can be enumerated as a vector of size $2 \lambda+1$, with $\mu \in [-\lambda, \lambda]$. For example, the dipole moment being a vector decomposes into a single irrep indexed by $\lambda = 1, \sigma=1$, whereas the polarizability tensor decomposes into irreps labelled by $(\lambda, \sigma)$ pairs corresponding to $(0,1), (1,1), (2,1)$.

These properties are modeled as additive quantities, i.e. as sums of atomic contributions %
and can be conveniently approximated in terms of atom-centered descriptors that exactly mirror the (improper) rotational nature of the target, combined with invariant weights,
\begin{equation}\label{eq:lambda-soap}
    y_{A}^{\sigma \lambda \mu} = \sum_{i\in A} y_{A_i}^{\sigma \lambda \mu} =  \sum_{i \in A} \mathbf{w}^{\sigma \lambda a_i}  \cdot  \boldsymbol{\xi}^{\sigma\lambda\mu}_{A_{i}}.
\end{equation}
$\boldsymbol{\xi}^{\sigma\lambda\mu}_{A_{i}}$ denotes the equivariant $\lambda$-SOAP~\cite{gris+18prl} descriptor for atom $i$ in structure $A$, where $a_i$ is the atomic species of $i$. The invariant model weights $\mbf{w}^{\sigma \lambda a}$ (invariance indicated by the absence of $\mu$ labels) are indexed by $\lambda$, $\sigma$, and $a$, to highlight that each target irrep can be learned by a distinct linear model. The dot-product in Eq.~\eqref{eq:lambda-soap} is taken over the feature dimension, which includes radial and angular components of the basis used to compute the $\lambda$-SOAP descriptor, and information about the chemical variability in the environment centered on $i$. 
In the following, we will use these kinds of symmetry-adapted regression models as examples of property models for dipoles and polarizabilities, analogous to the kernel models used in \texttt{MuML}~\cite{veit+20jcp} and \texttt{AlphaML}~\cite{wilk+19pnas}.
The choice of a simple ridge regression model is made in the same spirit as the restriction of the Hamiltonian model to a linear form -- keeping a minimalistic form of ML with a restricted design space to focus on the effect of the ``QM-facing'' part of the model architecture.

\subsection{ML models of effective single-particle Hamiltonians} 

In contrast to global properties such as dipoles and polarizabilities, which are modeled as a sum of atom-centered contributions but are trained against references computed for the entire molecular structure, the Hamiltonian matrix elements depend on specific pairs of orbitals involved in the interaction. 
When these orbitals are centered on atoms, as is the case for localized AO bases, the Hamiltonian matrix elements can be viewed as objects labeled by pairs of atoms, as well as multiple quantum numbers, namely the radial ($n$) and the angular ($l, m$) quantum numbers characterizing each AO.
These angular functions are typically chosen to be (real) spherical harmonics, and determine the equivariant behavior of the matrix elements under rotations and inversions. 
The non-equivalence of matrix elements under the exchange of atom labels, while keeping the orbitals fixed, also makes them equivariant under permutations of the atom labels~\cite{niga+22jcp}. 
 
In the same spirit of equivariant models of global properties, we describe the rotational behavior of $\mathbf{H}$ by transforming each pair of angular functions into irreps of O(3). For each pair of angular quantum numbers $(l, l')$ associated with the radial labels $n$ and $n'$ for atoms $i$ and $j$, we couple the angular functions using Clebsch–Gordan coefficients to obtain equivariant outputs indexed by $\lambda \in [\,|l-l'|,\,l+l']$,

\begin{align}
\label{eq:ham-block}
    \mbf{H}^{\mbf{p} \sigma \lambda \mu}_{A_{ij}} = \sum_{m m'} \cg{\lm}{l'm'}{\glm} \mbf{H}^{\mbf{p} m m'}_{A_{ij}},
\end{align}
where we use the shorthand $\mbf{p}=(n,l,n',l')$ to denote the combined set of indices for the angular and radial basis functions, as well as the chemical species of the atoms, and $\cg{\lm}{l'm'}{\glm}$ are the Clebsch-Gordan coefficients.

To address the two-centered nature of the matrix elements and construct a model akin to Eq.~\eqref{eq:lambda-soap}, we extended atom-centered descriptors in Ref.~\onlinecite{niga+22jcp} to a framework capable of describing multiple atomic centers and their connectivities, giving rise to the equivariant pair descriptor $\boldsymbol{\xi}^{\sigma\lambda\mu}_{A_{ij}}$, which simultaneously characterizes the environments of atoms $i$ and $j$ in structure $A$.
The features are built as a symmetrized product of density expansion coefficients for atomic pairs and atom-centered neighbor densities, and were also used in Ref.~\citenum{cignoni2024electronic} (see the SI for a full discussion of the features and their hyperparameters). 
Here, $\lambda\mu$ denote the rotational symmetry, and $\sigma$ indicates inversion parity, as in Eq.~\eqref{eq:lambda-soap}.  
Each Hamiltonian block ~\eqref{eq:ham-block} can be modeled separately, for instance, through a linear layer,

\begin{equation}\label{eq:direct-hamiltonian-model}
    \mbf{H}_{A_{ij}}^{\mbf{p}\sigma\lambda\mu} = \mbf{w}^{\mbf{p}\sigma\lambda} \boldsymbol{\xi}^{\sigma\lambda\mu}_{A_{ij}} +\delta_{\lambda 0} b^{\mathbf{p}},
\end{equation}
where $\mbf{w}^{p\sigma\lambda}$ is an invariant weight and the intercept $\delta_{\lambda 0} b^{\mathbf{p}}$ is nonzero only for invariant blocks, to maintain equivariance. 

\subsection{Electronic Hamiltonians as elements in an ML architecture}

Matrix elements of the effective one-electron Hamiltonian have not only served as targets for machine learning but also as inputs to models that have successfully predicted a wide range of molecular properties~\cite{fabrizio2021spa,welborn2018transferability,qiao2020orbnet}.
Hybrid ML-QM models treat $\mathbf{H}$ in yet another way, using an ML prediction of the Hamiltonian as an intermediate component in a modular pipeline that computes the desired chemical properties, given an input molecular configuration~\cite{cignoni2024electronic}. 
Rather than explicitly targeting the matrix elements obtained from a QM calculation, the Hamiltonian is learned implicitly, as an intermediate layer used to predict several physical properties accessible in a quantum chemistry calculation, which are used as the model targets. The implicit Hamiltonian does not need to be expressed in the same basis as used for calculating reference values of these target properties, enabling the emulation of large basis-set calculations with a significantly simpler and smaller model. In Ref.~\onlinecite{cignoni2024electronic}, this procedure demonstrated promising results in both extrapolating to diverse structures, including molecules much larger than the training set, and generalizing to observables not optimized during training, such as molecular excitations.

To extend the capabilities of these models beyond direct diagonalization, and facilitate the computation of observables that require non-trivial manipulations of the Hamiltonian, we interface our ML model with \textsc{PySCFAD}, an electronic structure code that supports automatic differentiation~\cite{zhang2022differentiable}, as shown schematically in Fig.~\ref{fig:model-schematic}. This approach delegates the physics-based operations required to obtain properties from the ML-predicted Hamiltonians to \textsc{PySCFAD}.
The availability of such a modular framework, that provides automatic differentiation of the target properties with respect to the intermediate Hamiltonian (and, by back-propagation, with respect to the parameters of its ML representation) opens up the possibility of investigating how the accuracy and transferability of these hybrid models depend on the choice of learning targets, and more generally on the details of the part of the model architecture that is explicitly built to mimic physics-based manipulations of $\mathbf{H}$.

\section{Results}

\subsection{Training and test set construction}

For training, we use a subset of 700 molecules from the QM7b dataset~\cite{blum,montavon2013machine}, containing only molecules composed of C, H, N, and O atoms. These molecules were selected from a total of 9,000 using farthest point sampling on three body atom-centered descriptors (SOAP~\cite{bart+13prb}) to maximize structural diversity. Details of sample selection, reference electronic structure calculations, and generation of the dataset are provided in Sec.~S1 in the SI. 
We use calculations performed with two different basis sets -- a minimal STO-3G basis, and a larger def2-TZVP basis.
The accuracy of the model is evaluated on two separate test sets, the first consisting of 100 QM7 molecules, and the second composed of 200 QM9 molecules~\cite{ramakrishnan2014quantum, ruddig}, both spanning the same compositional space. In addition, we assess performance on a few targeted benchmarks as discussed below. 

\subsection{The design space of effective Hamiltonian models}

When using a model that explicitly learns the matrix elements of a target $\mathbf{H}$ in a given AO basis set, one does not need to separately learn the AO representation of the overlap matrix, or that of operators needed to compute other properties (e.g. the position operator), as they can be computed inexpensively in the same basis.
In contrast, when using an indirect model that learns one or more derived properties -- and particularly, when targeting property values computed with a different basis -- the intermediate Hamiltonian learned within the model does not correspond to a well-defined basis (as discussed further below). 
Consequently, the overlap matrix and the AO representation of various operators must also be redefined for consistency. These could either be considered as additional components within the model, having the same size and symmetries compatible with the parameterization of $\mathbf{H}$, but could originate from any compatible basis set, or learned separately. 
We experimented with learnable overlap and operators, but found that this approach led to model instabilities as additional constraints (e.g. enforcing the overlap to be positive-definite) were necessary to maintain physical consistency and numerical stability. 

In the following, unless stated otherwise, we use a linear model, and the Slater-type orbital (STO)-3G basis featuring a minimal number of AOs on each atom as the model basis.
To compute functions of the Hamiltonian, we use the exact STO-3G representation of the overlap as well as AO representation of all operators computed on this basis. To improve the model convergence, we initialize the model weights to those obtained from a symmetry-adapted ridge regression~\eqref{eq:direct-hamiltonian-model} targeting the self-consistent single-particle Hamiltonian matrix from a minimal basis QM calculation.
We denote this initial model by $(\mathbf{H}_{\mathrm{STO\mbox{-}3G}})$. The model weights are subsequently refined using stochastic gradient descent on specific target properties derived from the predicted $\mathbf{H}$.
More details about the training procedures are provided in the Supplementary Information (SI).

\begin{figure*}
    \centering
    \includegraphics[width=0.9\linewidth]{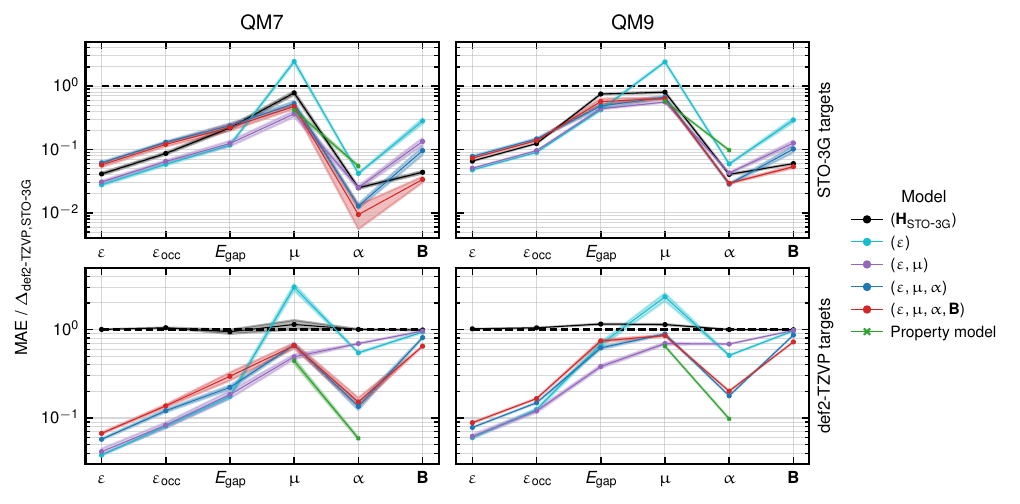}
    \caption{Comparison of model accuracies on QM7 and QM9 test datasets for various molecular properties (listed on the x-axis) computed using STO-3G (top) and def2-TZVP (bottom) basis sets. Each plot shows the ratio of MAEs from the pre-fitted ridge regression model $(\mathbf{H}_{\mathrm{STO\mbox{-}3G}})$ and different indirect models (denoted by the properties they are optimized on in parentheses), to the basis set error ($\Delta_{\mathrm{def2\mbox{-}TZVP,STO\mbox{-}3G}}$) which is the mean absolute error between the reference values computed in def2-TZVP basis and the STO-3G basis.
    This basis-set error serves as a baseline to compare different models and is indicated by the black, dashed line at $y=1$, and corresponds to (for QM7 and QM9, respectively): 4.99 eV and 4.87 eV for all MO energy levels ($\varepsilon$) 2.86 eV and 3.04 eV for the occupied energy levels $\epsilon_\text{occ}$, 
    2.60 eV and 1.36 eV for the HOMO-LUMO gap $E_\mathrm{gap}$, 0.017 a.u and 0.023 a.u for the dipole moment $\bm{\mu}$, 5.66 a.u and 5.82 a.u for the polarizability $\bm{\alpha}$ and 5.93 a.u and 7.26 for the Mayer bond order $\mathbf{B}$.}
    \label{fig:model comparison}
\end{figure*}

We assess the performance of two classes of indirect Hamiltonian models as additional target properties are gradually introduced. The first class, referred to as \emph{simple} indirect models, implicitly represents the Hamiltonian in the same AO basis as used for target calculations. The second class, in contrast, comprises  \emph{upscaled} indirect models that target properties from a larger, more converged AO basis (and potentially even a different level of theory), while the predicted Hamiltonians are expressed in the smaller model basis. In the following, we denote the large basis as LB. 
For each model class, we progressively include the following target properties during optimization: MO energies $(\bm{\varepsilon})$, molecular dipole moments $(\bm{\mu})$, molecular polarizabilities $(\bm{\alpha})$, and Mayer bond order $(\mathbf{B})$~\cite{mayer}. The Mayer bond order is defined as

\begin{equation}
    B_{ij} = \sum_{\eta \in i} \sum_{\eta' \in j} (\bm{\rho} \mathbf{S})_{\eta\eta'} (\bm{\rho}\mathbf{S})_{\eta'\eta},
\end{equation}
where $\eta$ and $\eta'$ label AOs centered on atoms $i$ and $j$, respectively. It is representative of the localization of the density matrix in the AO basis used for the target calculations, while remaining independent of the choice of the specific AO labels. In the following, these fine-tuned models are denoted by the properties on which they are optimized, for instance, $(\bm{\varepsilon})$ indicates a model trained solely on MO energies, while $(\bm{\varepsilon},\bm{\mu})$ refers to the one trained on both MO energies and dipole moments, and so on.

Fig.~\ref{fig:model comparison} illustrates the relative performance of different models on the QM7 and QM9 test datasets. The top row shows results for the simple indirect models, whereas the bottom row corresponds to upscaled indirect models where the targets are computed using LB, which in this case, is the def2-TZVP basis. We evaluate the performance of different models across the prediction of several properties, namely, MO energies, occupied-state MO energies ($\bm{\varepsilon}_{\mathrm{occ}}$), dipole moments, polarizabilities, HOMO–LUMO gaps ($E_\mathrm{gap}$), and Mayer bond orders, as listed on the $x$-axis. The normalized mean absolute error (MAE) of each model is reflected on the $y$-axis. Note that these are normalized by the \emph{basis set error}, ($\Delta_{\mathrm{def2\mbox{-}TZVP,STO\mbox{-}3G}}$), which is defined as the MAE between the reference observables computed in the def2-TZVP and the STO-3G basis, such that $y=1$ indicates the magnitude of error associated with the convergence of the basis set.
As expected, for both the QM7 and QM9 test sets, the prediction accuracy of a property is improved if it is explicitly included in the optimization. However, imposing additional constraints may impact accuracy due to the redistribution of the model flexibility and available training data over several optimization tasks.
Overall, the results indicate that models optimizing all ($\varepsilon,\bm{\mu},\bm{\alpha},\bm{B}$) are the most robust. 
With the exception of the prediction of dipole moments from $(\bm{\varepsilon})$ models, almost all model errors are between one and two orders of magnitude smaller than the basis set error. 

Simple indirect models incur much lower MAEs as their implicit Hamiltonian representation is consistent with that of the target, and the model does not have to compensate for basis set convergence -- especially for properties such as $\alpha$ and $B$ that exhibit a strong basis-set sensitivity.  
However, upscaled indirect models need to effectively extrapolate beyond the STO-3G basis to match the def2-TZVP targets, leading to larger ML errors. This implies that basis set incompleteness is a dominant source of error and that ML struggles to fully correct for it. 
We also compared these models with property-specific surrogate models, analogous to \texttt{MuML} and \texttt{AlphaML} ~\eqref{eq:lambda-soap}, trained on the same training dataset as the indirect models to predict dipole moments and polarizability computed in the two different bases. Simple indirect models improve the prediction accuracy of polarizability in the STO-3G basis compared to the property models in the same basis.
However, the property models for both dipoles and polarizabilities in the def2-TZVP basis consistently exhibit lower errors than the indirect upscaled models. Thus, even though it is possible to reproduce molecular properties computed with a large basis using the indirect upscaled model with a minimal basis representation of the Hamiltonian, these predictions may be inherently limited by the representability of the model basis. The size of the model Hamiltonian must, therefore, be considered a part of the model hyperparameters.

Another important factor affecting the accuracy of response properties is the accuracy of the virtual MOs. Small basis sets often fail to adequately represent the virtual states due to an insufficient availability of polarization functions. According to standard linear response theory~\cite{baroni2001phonons}, the first-order correction to the density matrix under an external electric field $\bm{\mathcal{E}}$ is given by $\Delta \rho_{\eta \nu} = \bm{\mathcal{E}} \cdot \bm{\chi}_{\eta\nu}$, where the response vector $\bm{\chi}_{\eta\nu}$ is defined by

\begin{align}
    \bm{\chi}_{\eta \nu} = 2 \sum_{r \in \mathrm{occ}} \sum_{s \in \mathrm{virt}} \frac{\mathbf{x}_{rs}}{\varepsilon_r - \varepsilon_s} \left(C_{\eta s} C_{\nu r} + C_{\eta r} C_{\nu s}\right).
\end{align}

Here, $\eta,\nu$ index AOs, $r,s$ index MOs, $\mathbf{x}_{rs}$ denotes the matrix elements of the electronic position operator, and $\mathbf{C}$ is the MO coefficient matrix. As the denominator depends on virtual MO energies, enhancing their accuracy directly improves the prediction of polarizability and other linear response properties.

\begin{figure}[tb]
    \centering
    \includegraphics[width=\columnwidth]{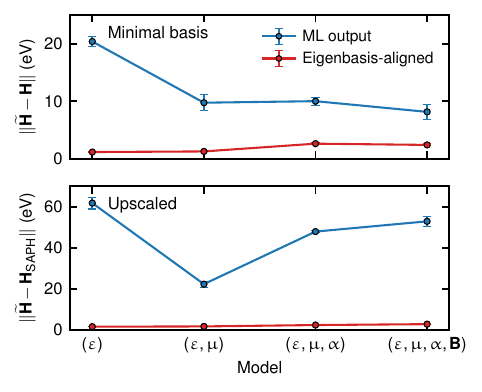}
    \caption{Norm of the deviations of the predicted Hamiltonians from their respective reference across the QM7 test set for simple indirect models (top) and upscaled indirect models (bottom). Blue lines show the deviations of ML predictions, while red lines show the deviations of eigenbasis-aligned predictions. Markers represented the values averaged over the dataset and three random train/test splits, and error bars indicate the standard deviations across the three splits of the dataset.}
    \label{fig:interpretability}
\end{figure}

\subsection{Interpreting effective Hamiltonians}

Even when an indirect model is represented in the same basis as its reference calculations (i.e. simple indirect models), the predicted $\widetilde{\mathbf{H}}$ at the end of the training need not exactly match the target $\mathbf{H}$. 
The matrices can differ because of a mismatch in the eigenvalues, or because of a different alignment of the eigenvectors.
In order to differentiate between the two effects, we define an \emph{eigenbasis alignment} (EA) operation
\begin{align}\label{eq:alignment}\widetilde{\mathbf{H}}_\mathrm{EA} = \mathbf{C} \widetilde{\mathbf{C}}^\dagger \widetilde{\mathbf{H}} \widetilde{\mathbf{C}} \mathbf{C}^\dagger.
\end{align}
$\widetilde{\mathbf{C}}$ and $\mathbf{C}$ are the eigenvector matrices of $\widetilde{\mathbf{H}}$ and $\mathbf{H}$, respectively (we consider for simplicity L\"owdin-orthogonalized matrices). 
If the predicted eigenspectrum exactly matches the reference one, $\widetilde{\mathbf{H}}$ and $\mathbf{H}$ are related by an orthogonal transformation, and $\widetilde{\mathbf{H}}_\mathrm{EA}=\mathbf{H}$. 
In the general case, where predictions have some error, eigenbasis alignment can still be considered the best orthogonal transformation relating $\widetilde{\mathbf{H}}$ and $\mathbf{H}$, in the sense that the discrepancy between $\mathbf{H}$ and $\widetilde{\mathbf{H}}_\mathrm{EA}$ correlates with the discrepancy between the two eigenspectra. In other words, we can consider the difference between $\|\mathbf{H}-\widetilde{\mathbf{H}}\|$ and $\|\mathbf{H}-\widetilde{\mathbf{H}}_{\mathrm{EA}}\|$ as a measure of how much indirect training rotates the representation of the predicted Hamiltonians with respect to the original basis set. 

For upscaled models, where the target properties are computed using LB than the one used in the model, eigenbasis-aligned predictions can be compared with symmetry-adapted projected Hamiltonians (SAPHs)~\cite{niga+22jcp} defined as

\begin{align}\label{eq:saph}
    \mathbf{H}_{\mathrm{SAPH}} = \mathbf{C} \overline{\mathbf{C}}_{\mathrm{LB}}^\dagger \mathbf{H}_{\mathrm{LB}} \overline{\mathbf{C}}_{\mathrm{LB}} \mathbf{C}^\dagger,
\end{align}
where $\overline{\mathbf{C}}_{\mathrm{LB}}$ is the submatrix of eigenvectors of $\mathbf{H}_{\mathrm{LB}}$ corresponding to the number of molecular orbitals described by the model's basis, i.e. the STO-3G basis, and $\mathbf{C}$ is the same as in Eq.~\eqref{eq:alignment}. We compare these quantities in Fig.~\ref{fig:interpretability}, which illustrates the deviations of $\widetilde{\mathbf{H}}$ and $\widetilde{\mathbf{H}}_{EA}$ from the reference Hamiltonian $\mathbf{H}$. As expected, the deviations after EA (shown by the red curve) are significantly smaller than those of the original ML outputs (shown in blue). For minimal basis targets (top panel), increasing the number of constraints directly correlates with decreased deviations from the reference in the case of the original ML outputs, revealing some correlation between larger number of constraints and similarity between $\widetilde{\mathbf{H}}$ and $\widetilde{\mathbf{H}}^{\mathrm{EA}}$. Given the large number of target observables required to fully constrain the flexibility of representation and the fact that we only have four data points, the seemingly decreasing behavior might be fortuitous. 
In fact, the same trend does not hold for upscaled models (bottom panel).
The rotation of the indirect Hamiltonian relative to the reference small basis might help, especially in the case of upscaled predictions, reduce the effect of having operators computed in a fixed (and different) basis.

\begin{figure}[tb]
    \centering
    \includegraphics[width=\linewidth]{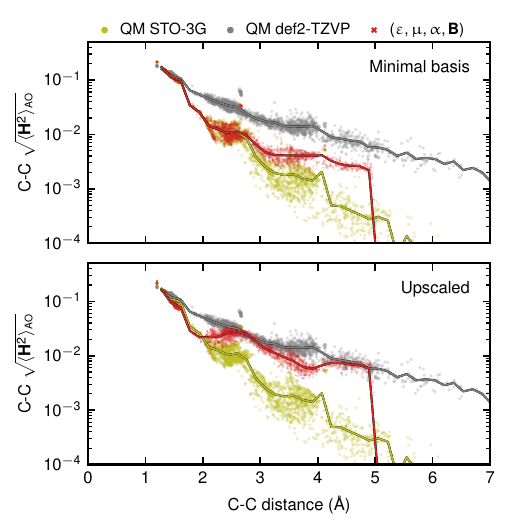}
    \caption{Decay of the carbon-carbon interaction terms in the QM9 test set as predicted from (a) simple indirect (targeting  STO-3G observables) and (b) upscaled indirect (targeting def2-TZVP observables) models, both trained with minimal-basis model architectures. When upscaling the model to predict large-basis properties, the effective Hamiltonian adapts its behavior to match the natural scale of the reference-calculation Hamiltonian. Solid lines are moving averages of the data points.}
    \label{fig:h decay}
\end{figure}

We further analyze the effective Hamiltonians in terms of the decay of their matrix elements as a function of the distance between the two interacting atoms. In Fig.~\ref{fig:h decay} we examine carbon-carbon interactions more closely. %
The left panel shows how $\widetilde{\mathbf{H}}$ from a simple indirect model ($\varepsilon,\bm{\mu}, \bm{\alpha},\mathbf{B}$), denoted by the red dots, adapts to the reference STO-3G matrix elements while on the right panel the $\widetilde{\mathbf{H}}$ from the corresponding upscaled indirect model adapts to the scale and decay characteristics of the reference def2-TZVP matrix, despite being represented in a smaller basis. This highlights that the indirect Hamiltonian models implicitly capture characteristic spatial patterns and magnitudes of Hamiltonian elements specific to the target basis, providing physically consistent effective Hamiltonians even when trained indirectly on derived properties.

\subsection{Extrapolative case studies}\label{sec:showcase}

\begin{figure}[tb]
    \centering
    \includegraphics[width=\linewidth]{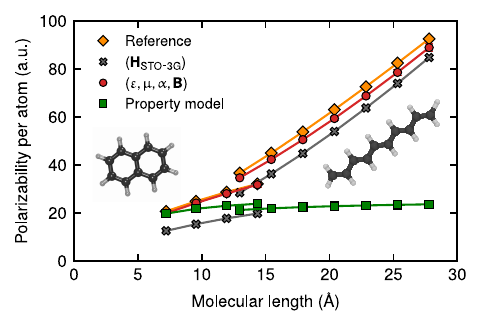}
    \caption{Norm of the polarizability per atom for increasing chain lengths of polyalkenes and polyacenes. Orange diamonds indicate the reference values, gray crosses show predictions from our Hamiltonian model before fine-tuning (\mbf{H}$_{\mathrm{STO\mbox{-}3G}}$), red circles represent predictions from the upscaled indirect Hamiltonian model ($\varepsilon, \bm{\mu}, \bm{\alpha}, \mathbf{B}$), and green squares correspond to the property-specific model ~\eqref{eq:lambda-soap} resembling \texttt{AlphaML}. Error bars indicate standard deviations over three independent random test/train splits. Even the Hamiltonian model prefitted to STO-3G qualitatively captures the correct scaling with system size, while the property-specific model does not. After fine-tuning, the Hamiltonian model predictions become quantitatively accurate. Example molecular structures from the dataset are shown in the insets.}
    \label{fig:polarizability alphaml}
\end{figure}

\begin{figure}[tb]
    \centering
    \includegraphics[width=\linewidth]{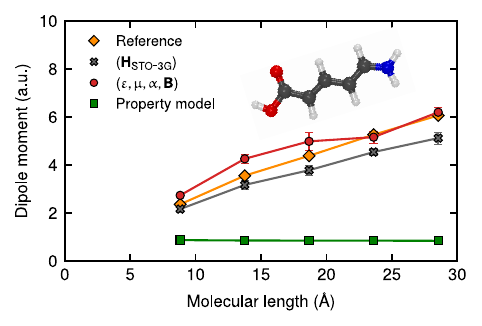}
    \caption{Norm of the dipole moment for a series of polyenoic amino acids with increasing chain length. Orange diamonds indicate the reference values, gray crosses show predictions from the pre-fitted Hamiltonian-based model \mbf{H}$_{\mathrm{STO\mbox{-}3G}}$, red circles mark predictions from the Hamiltonian-based model after fine-tuning with loss function $\mathcal{L}=\mathrm{MSE}_{\bm{\varepsilon}, \bm{\mu}, \bm{\alpha}, \mathbf{B}}$, and green squares correspond to the property-specific model ~\eqref{eq:lambda-soap} analogous to \texttt{MuML}. Error bars indicate standard deviations over three independent random test/train splits. Even without fine-tuning, the Hamiltonian-based approach qualitatively captures the correct scaling with system size, while the property-specific model does not. After fine-tuning, the Hamiltonian model predictions become quantitatively accurate. An example molecular structure from the dataset is shown in the inset.}
    \label{fig:dipole muml}
\end{figure}

In the extrapolative tests on QM9, shown in Fig.~\ref{fig:model comparison}, the degradation in model accuracy suffered by the Hamiltonian-based models is similar to that suffered by property models that directly estimate dipole moment and polarizability. 
Refs.~\onlinecite{wilk+19pnas,veit+20jcp}, however, discuss specific test cases that lead to \emph{qualitative} failures in the property models, linked to their local, atom-centered nature.
The first of such test datasets involves the polarizability of long-chain polyalkenes and polyacenes, while the second consists of a series of polyenoic amino acids featuring an amine and a carboxylic acid group linked by a polyacetylene backbone.  
In the first series, the polarizability per atom increases with chain length owing to the progressive reduction of HOMO-LUMO gap, and the resulting delocalization of electrons. 
Similarly, in the second dataset, the molecular dipole moments grow with system size as a result of the delocalization of charges over the entire molecule. %
Traditional surrogate property models, which map molecular geometry to physical observables via Eq.~\eqref{eq:lambda-soap}, fail to reproduce the correct scaling behavior due to their strictly local architectures. 
These models describe atomic environments up to distances much smaller than the characteristic length scales at which these phenomena occur, and the training set contains only small molecules.

In contrast, model architectures incorporating an effective Hamiltonian replicate the correct physical behavior. As shown in Figs.~\ref{fig:polarizability alphaml} and~\ref{fig:dipole muml}, even a model fitted to \mbf{H}$_{\mathrm{STO\mbox{-}3G}}$ that we use as an initialization for all other models, qualitatively captures the scaling laws of the properties with respect to molecular length, and quantitative accuracy is achieved by further fine-tuning. Even though the Hamiltonian matrix elements are also predicted with a strictly local model, the physics-based manipulations, most importantly the diagonalization of the effective Hamiltonian, introduce the non-local physics that is necessary to reproduce the correct trends.
These observations concretely exemplify the superior transferability of indirect Hamiltonian models compared to property-specific models.

\section{Condensed-phase Hamiltonians}

\begin{figure}[tb]
    \centering
    \includegraphics[width=\linewidth]{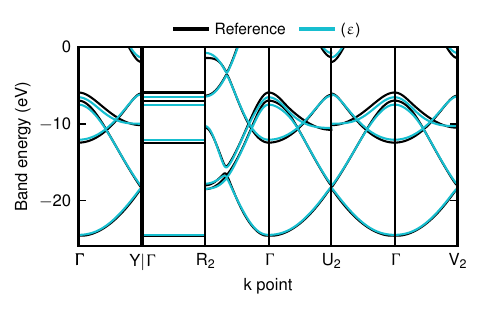}
    \caption{Comparison of reference and predicted bands for a hold-out structure from the dataset. The predictions are from an upscaled SVP-basis model targeting the correction to the $(\bm{\varepsilon})$ bands computed with a more converged DZVP basis.}
    \label{fig:graphene}
\end{figure}

It is important to stress that, even though we have exclusively presented molecular examples thus far, the indirect learning strategy presented in this work can also be readily applied to the condensed phase. Extended systems are routinely treated under the Born-Von K\'arm\'an periodic boundary conditions (PBC), repeating the crystal unit cell several times along each lattice vector, and finally imposing the periodicity of the electronic wavefunction on this large (super-)cell. Although accounting for periodicity is more natural and convenient in a reciprocal space representation, the structural inputs parameterizing the Hamiltonian describe real-space geometries. 
Thus, the predicted real-space Hamiltonians corresponding to each translation $\mathbf{R}_\mathbf{t}$ of the unit cell can be transformed to the reciprocal space through the Bloch summation, %
\begin{align}
    \mathbf{H}(\mathbf{k}) = \sum_{\mathbf{t}} \mathbf{H}(\mathbf{t}) e^{i \mathbf{k} \cdot  \mathbf{R}_{\mathbf{t}}}.
\end{align}
$\mathbf{R}_\mathbf{t}$ denotes the Bravais lattice vector and is labeled by the three integers $\mathbf{t}=(t_x, t_y, t_z)$ indicating the number of unit-cell repetitions in each Cartesian direction. $\mathbf{H}(\mathbf{k})$ is then diagonalized to obtain a set of band energies, $\{\varepsilon_{n\mathbf{k}}\}$, labeled by a band index $n$ and a Brillouin zone vector, $\mathbf{k}$, that can be indirectly learned on an upscaled basis similar to the molecular case. 
As a simple example, we consider a small dataset of 23 structures subsampled from the graphene dataset from Ref.~\onlinecite{li2022deeph}. Instead of directly learning the large basis Hamiltonian (in this case, the reference calculations are in the DZVP (double-zeta valence polarized) basis) we construct a $(\bm{\varepsilon})$ model trained on the band energies up to a few eV above the Fermi level (more specifically, all the occupied states plus one empty state per C atom),  using a calculation in the minimal SZV (single zeta valence) basis as baseline. 
For these baseline DFT calculations, we use a G\"odecker-Teter-Hutter (GTH) pseudopotential~\cite{goedecker1996separable}, and note that the basis functions are more spatially delocalized than those used for all-electron calculations. To facilitate modeling with (relatively) local descriptors, we target the difference between SZV and DZVP results rather than pre-training the model on the self-consistent Hamiltonians in the smaller-basis as in the molecular case. The ML model predicts the correction to SZV calculations necessary to achieve large-basis accuracy at inference.  
The model achieves an MAE of 252\,meV  on the occupied band energies. The predicted bands for a hold-out test structure are compared against the reference in Fig.~\ref{fig:graphene}. 

\section{Conclusions}
Machine learning has been transformative for chemical sciences, offering faster and more accurate predictions of molecular properties and significantly expanding the scope of computational chemistry. In most applications, machine learning models are used as surrogates for quantum mechanical calculations. Over the past few years, there has been increasing interest in directly modeling elements of a QM calculation, such as the effective single-particle Hamiltonians, from which molecular properties can be subsequently derived. Previous works in this context have focused on targeting Hamiltonian matrix representations on a specific basis, or learning a reduced effective representation which reproduces observables from a calculation performed on another, larger basis~\cite{schu+19nc, li2022deeph, gong2023natcom, zhang2022equivariant, cignoni2024electronic}.

Whenever the Hamiltonian is not an explicit target, the way it is parameterized and the targets chosen for training become part of the model architecture, introducing quantum chemistry considerations into the modeling design space.
To facilitate the exploration of these possibilities, we introduced a framework that seamlessly integrates the ML predictions of effective one-electron Hamiltonians with \textsc{PySCFAD}, which is a differentiable quantum chemistry code. 
This allows us to move beyond targeting the matrix elements in a fixed basis, and instead treat the Hamiltonian as an intermediate model layer that can be optimized on multiple observables computed from it through differentiable analytical operations. 

A key advantage of this framework is its flexibility, allowing it to simultaneously target numerous observables which need not be limited to those derived from a QM calculation performed in the same basis as the one used to represent the implicit model Hamiltonian.
We demonstrated the capabilities of the framework by predicting molecular orbital energies, dipole moments, and polarizabilities for subsets of the QM7 and QM9 datasets, computed using both STO-3G and def2-TZVP basis sets, while restricting the model Hamiltonian to have the size and symmetries compatible with STO-3G. 
In both cases, we gradually constrain the model on energies, dipoles, polarizabilities and Mayer bond order. The addition of each constraint improves the prediction of the corresponding property at the expense of a possible decrease in accuracy for the remaining ones, as the model expressivity and available training information is redistributed among the different targets. 
During training, the intermediate Hamiltonian is free to deviate from the reference STO-3G basis to a form that is more suited for training, and we observe that, when targeting def2-TZVP-computed properties, the implicitly learnt Hamiltonian tends to adapt to the slower decay of the LB. 
The indirect prediction of quantum mechanical properties shows good accuracy and transferability to larger molecules despite the fact that we restricted the structural representation of the input molecules to fixed, low-body-order descriptors and the mapping of the matrix elements to a linear form. For structural properties such as polarization and polarizability, the Hamiltonian-based predictions are comparable to those of bespoke models targeting the final quantity directly. 
Contrary to the latter, our approach correctly captures the qualitative trends in predictions for the response properties of complex molecules, such as polyenes, polyacenes, and polyenoic amino acids, where traditional property-specific models have been shown to struggle or fail entirely, as they do not correctly account for electron delocalization.
This approach extends beyond molecular systems as exemplified by the predictions of DZVP band structure of graphene from a model implicitly restricted to an SZV representation of the Hamiltonian, that are within 3\% of the reference.

This work underscores the potential of a hybrid QM-ML paradigm in enhancing the transferability of predictive models across geometric complexity, QM details (basis set, level of theory), and physical observables. Although we restricted ourselves to simple ML models to emphasize the QM design choices, the framework we propose here is equally applicable to more sophisticated ML architectures, which could further improve model accuracy and propel predictive modeling in complex systems.
The elements of the architecture that are most closely related to the QM workflow -- most notably the size and assumed symmetry of the basis -- are equally important. Especially when targeting properties from a more converged basis, the constraints or observables that are explicitly included in the training affect (but do not determine entirely) the alignment between the effective basis predicted by the model and that used to represent the operators that underlie those properties. 
For example, the use of a minimal basis representation is only partly compensated by the ability of the model to learn an effective description of the more converged basis. In other terms, restrictions on the intermediate model basis can reduce the expressive power of the model, especially for excited states and properties such as the polarizability that are strongly dependent on them. 
Overall, we suggest that when applying ML to quantum chemical calculations, one has to take a holistic approach in which geometric features, model architecture, and QM approximations are optimized together. A higher degree of integration between the ML and QM software stack -- as we implement here exploiting the autodifferentiability of \textsc{PySCFAD} -- helps explore the design space of these hybrid models, exploit synergies between the different parts of the calculation and ultimately improve the accuracy, transferability and computational efficiency of the predictions.

\begin{acknowledgments}
MC, DS, JN, and SS acknowledge funding from the European Research Council (ERC) under the research and innovation program (Grant Agreement No. 101001890-FIAMMA), an industrial grant from Samsung, and the NCCR  MARVEL, funded by the Swiss National Science Foundation (SNSF, grant number 205602).
PP acknowledges funding from Samsung, HT acknowledges funding from the Walter Benjamin program (project number 534232519) of the German Research Foundation (DFG).
GKC and XZ acknowledge support from the US Department of Energy, Office of Science, Basic Energy Sciences, Chemical Sciences, Geosciences, and
Biosciences Division under Triad National Security, LLC
(``Triad'') contract Grant 89233218CNA000001. 
\end{acknowledgments}
\textbf{Conflict of interest statement}\\
The authors have no competing interests to declare.\\
\textbf{Author contributions}
MC and GKC designed the research strategy; DS, JN, PP and SS performed simulations and prepared figures; JN and PP developed the ML software infrastructure, with contributions from DS, SS and HT; XZ and GKC developed the \textsc{PySCFAD} interface; JN, PP and DS wrote the first version of the article; all authors commented on the manuscript and contributed to the final version.\\
\textbf{Data and materials availability}\\All the software used in this study is publicly available at Ref.~\citenum{mlelec-repo}.
Datasets comprising molecular configurations and reference electronic-structure properties will be made available upon publication.
\end{document}